\documentclass{llncs}

\usepackage{booktabs}

\usepackage{url}
\newcommand{\footurl}[1]{\footnote{\url{#1}}}

\usepackage{amssymb}
\newcommand{\N}{\mathbb{N}}

\title{Recommendations to OSCE/ODIHR (on how to give better recommendations for Internet voting)}

\author{Jan Willemson}

\institute{Cybernetica, Narva mnt 20, Tartu 51009, Estonia}

\date{}

\begin{document}
\maketitle

\begin{abstract}
    This paper takes a critical look at the recommendations OSCE/ODIHR has given for the Estonian Internet voting over the 20 years it has been running. We present examples of recommendations that can not be fulfilled at all, but also examples where fulfilling a recommendation requires a non-trivial trade-off, potentially weakening the system in some other respect. In such cases OSCE/ODIHR should take an explicit position which trade-off it recommends. We also look at the development of the recommendation to introduce end-to-end verifiability. In this case we expect OSCE/ODIHR to define what it exactly means by this property, as well as to give explicit criteria to determine whether and to which extent end-to-end verifiability has been achieved.
\end{abstract}

\section{Introduction}
\label{sec:introduction}

Organization for Security and Co-operation in Europe (OSCE) through its Office for Democratic Institutions and Human Rights (ODIHR)\footurl{https://www.osce.org/odihr}\footnote{Even though the full formal acronym of the organization is OSCE/ODIHR, we will just use ODIHR in this paper for brevity.} is one of the major international organizations engaged in election observation. Targeting primarily the electoral processes of the 57 OSCE member states\footnote{\url{https://www.osce.org/participating-states}; despite the organization's name referring to Europe, there are also several member states from Asia and North America.}, ODIHR's contribution to setting the standards for international election observation reaches far beyond OSCE. This is achieved thanks to publishing very detailed observation reports and an excellent series of election handbooks.\footurl{https://www.osce.org/odihr/elections/handbooks}

ODIHR reports cover all the major aspects of elections from candidate registration, minority issues, campaign financing and media to election administration and assessment of technical details of voting. The latter also covers electronic methods including voting machines and vote casting over Internet. 

Several OSCE member states (including Estonia, France, Norway, Russia, Switzerland and United States of America) have allowed Internet voting at various times and circumstances. Out of these countries, Estonia has enabled casting votes via Internet in legally binding elections continuously the longest, since 2005. (For a historical overview and some technical details of Estonian Internet voting we refer the reader to Ehin \emph{et al.}~\cite{EHIN2022101718}.) By 2025, Estonia has had 5  municipal, 5 parliamentary and 4 European Parliament elections allowing to cast the votes over Internet.\footurl{https://www.valimised.ee/en/archive/statistics-about-internet-voting-estonia}

Since ODIHR observes parliamentary and presidential elections, its missions have produced five final reports covering Internet voting in Estonia.\footnote{In Estonia, presidential elections are not national, but are held in the parliament or by an extended electoral body comprised of members of the parliament and representatives of the municipal councils.} These reports provide a valuable insight into the development of Internet voting in general, extending beyond the mere Estonian case study. 

This paper looks back at the 20 years of Internet voting in Estonia through the lens of ODIHR reports. We review some of the (more controversial) recommendations given in the reports, their context and evolution over the years. 

\section{Background}

ODIHR has produced five final reports on Estonian Internet-enabled parliamentary elections over the years\footnote{ODIHR also produces other kinds of documents, for example Needs Assessment Mission reports and preliminary versions of the final reports. However, the official recommendations are given in the final versions, and thus we concentrate on them.}, respectively from 2007~\cite{ODIHR2007}, 2011~\cite{ODIHR2011}, 2015~\cite{ODIHR2015}, 2019~\cite{ODIHR2019} and 2023~\cite{ODIHR2023}. These reports typically start from describing the general operational principles of the system, followed by the observations made by the rapporteurs, and recommendations. 

Even through ODIHR has no legal authority over the country's electoral system, the recommendations are still considered as guidelines for improvement. Systematic failure to meet some of the recommendations may shape the attitude of international community regarding the state of democracy in the country. This is why authorities generally try to act upon the ODIHR guidelines. In the countries where democratic traditions are not yet well established, these recommendations may be used e.g. by the civil society organizations to pressure the government to implement some changes.

On the other hand, potential reputation issues and pressure towards the authorities also mean that the recommendations should not be given lightly. It is easy to write a few sentences into a report, but if implementing them turns out to be unreasonable or even impossible, the recommendations may cause more harm than good. 

This has regrettably been the case in Estonia. Even if a recommendation is impossible to implement, failure to do so can be used by politically-motivated actors to spread distrust in the electoral system, including Internet voting. 

The goal of this paper is to give some recommendations to ODIHR on how to ensure that their recommendations are actually implementable and achieve a reasonable balance between the conflicting requirements of voting. 

Table~\ref{tab:ODIHRreports} presents a short summary of the five reports, showing how many pages were devoted to Internet voting, and how many formal recommendations were given.

\begin{table}[ht]
    \centering
    \caption{ODIHR reports about Estonian Internet voting}
    \label{tab:ODIHRreports}
    \begin{tabular}{ccc}
        \toprule    
        Year & Pages & Recommendations\\
        \midrule
        2007 & 12.5 & 9\\
        2011 & 7    & 13\\
        2015 & 4    & 8\\
        2019 & 2    & 5\\
        2023 & 3.5  & 6\\
        \bottomrule
    \end{tabular}
\end{table}

It is natural that in the first years more coverage was devoted to Internet voting as that was a very new thing, even at the international level. As we see, over the years 41 recommendations have been issued, with some of them being repeated or partially overlapping. The topics range from the proposals to update the legislation to very specific cryptographic ideas. Due to the page limit of the current paper we are unable to cover them all. Rather, we will be concentrating on the more interesting findings.

\section{Vote secrecy and coercion resistance}
\label{sec:votesecrecy}

Vote secrecy is one of the core requirements of democratic elections. Its goal is to ensure voting freedom by making it hard for the potential coercer to learn the voter's true preference. Conventional polling station paper voting attempts to achieve vote secrecy by enforcing the voters to fill in the ballots in private voting booths. This worked well in the 19th century Australia~\cite{brent2006australian}, and in the rest of the democratic world for the next 100\dots150 years. 

Due to the apparent success of secret ballot, this requirement is often also listed among desirata for other modes of voting (including remote ones) as well. This is also emphasized by ODIHR e.g. in their 2007 report~\cite{ODIHR2007} which states:
\begin{quote}
    ``Because the [Internet] voter is not voting in a supervised and controlled environment such as a polling station, it cannot be ensured that the voter is casting his/her vote in secret.''
\end{quote}

However, by the end of the 20th century, technological advancements in the recording equipment significantly weakened privacy guarantees of the polling booths~\cite{DBLP:conf/uss/Benaloh13}. It is possible to intercept the vote via various side channels~\cite{DBLP:conf/uss/CalandrinoC09,DBLP:journals/tissec/ToreiniSH17,DBLP:conf/voteid/KripsWV18,DBLP:conf/eurosp/KripsWV19} or simple video recording\footnote{\url{https://www.bbc.com/news/world-asia-56377642}, accessed January 30th, 2025} \footnote{\url{https://www.timesofisrael.com/hidden-cameras-in-arab-voting-booths-were-netanyahus-idea-tv-report/}, accessed January 30th, 2025} \footnote{\url{https://www.bbc.com/news/world-asia-68707488}, accessed January 30th, 2025}.

To compensate for the loss of privacy and potential coercion issues, Estonia offers the possibility to re-cast one's electronic vote in case the voter has felt coerced to vote against his/her true preference.\footnote{Most of the protective effect of re-voting actually comes from deterrence -- if the potential coercer knows that the voter can re-vote, he hopefully does not attempt to coerce the voter in the first place.} As a result, we can argue that the system becomes more secure against coercion than polling station paper voting. There it is enough for the coercer to observe (e.g. via video transmission) one act of paper voting and be sure that the vote will be counted as cast as there is no way to change a coerced paper vote afterwards. 

Of course, the information concerning which  electronic vote is the last one and will be tallied becomes sensitive as knowing it would help the coercer to achieve his goal. This is also noted by ODIHR in 2007~\cite{ODIHR2007}:
\begin{quote}
    ``However, the OSCE/ODIHR EAM noted that one technical aspect of the system undermines the objective of the recast possibility. Namely, the vote storage server records the time that each voter casts his/her last electronic vote. This log, which is available to political parties and observers, could potentially be misused to know whether a voter did in fact recast his/her vote electronically. \\[1ex]
    The OSCE/ODIHR EAM recommends that the NEC consider modifying the design of the internet voting system so that the time of voting is not recorded. In the interests of maintaining the transparency of the system, however, the log should continue to be available to observers.''
\end{quote}
This recommendation was repeated in 2011~\cite{ODIHR2011}.

Even though the concern of timed logs being sensitive is a valid one, just not recording the timestamps is not a viable solution. First of all, as noted above, these timestamps are needed to determine which one of the re-cast votes should be tallied. And second, the i-votes are digitally signed to provide both eligibility and integrity of the votes. The Digital Signature Act (later superseded by the EU eIDAS regulation) explicitly requires digital signatures to have timestamps to be able to determine whether the signature was given during the validity period of the signature key certificate.

\begin{quote}
    \emph{We recommend that, before giving recommendations, ODIHR analyses whether the prospective recommendation can be implemented in the first place, both from the technical and legal perspective.}
\end{quote}

Another potential issue with vote secrecy is presented in the 2019 report making use of the fact that the Estonian Internet voting protocol implements individual verifiability by making the encryption randomness available to the audit device via a QR-code~\cite{ODIHR2019}:

\begin{quote}
    ``A review of operational and technical frameworks by the ODIHR EET indicates that an internal attacker with privileged access to digital ballots could break the vote secrecy of any voter who published an image of the QR code online, even after the expiry of the code's validity. This contradicts national legislation and international standards pertaining to vote secrecy''
\end{quote}

It is noteworthy that an actual attack scenario is described here (which is something ODIHR does not do very often). Let us analyze this scenario in more detail. 

The attack involves two parties:
\begin{enumerate}
    \item  an internal attacker with privileged access to the digital ballots, and 
    \item  a voter willfully making his/her verification QR-code public.
\end{enumerate}
The attack has two steps:
\begin{enumerate}
    \item the voter publishes his/her individual verification QR-code online, and
    \item the insider uses the encryption randomness within the QR-code to open the digital ballot. 
\end{enumerate}

As a result, ODIHR recommends mitigating internal attacks, completely forgetting that the attack needs a voter willing to violate the secrecy of their own vote in the first place!

In the paper voting domain, the situation is more or less equivalent to a voter taking a selfie together with their filled in ballot (sometimes called \emph{stemfie} from the Dutch word \emph{stemmen}), and posting it online. No authority can essentially prevent it (even though some jurisdictions have attempted to outlaw stemfies, with varying degrees of success~\cite{koutsoulias2018ballot}). Internet voting is actually considerably more secure against this type of attack, as besides the voter willfully opening his/her vote it would also require a corrupt insider.

\begin{quote}
    \emph{We recommend that ODIHR gives a thorough consideration to the attacker models and required assumptions before publishing attack scenarios to justify the recommendations. Among other things, ODIHR should decide whether a voter interested in violating the secrecy of his/her own vote is considered an attacker or not. If yes, ODIHR should refer to this as a threat in all the reports about paper voting. Also, ODIHR should in this case present a clear strategy to fight against stemfies. If the voter willfully publishing their vote is not an attacker, ODIHR should retract the above-cited paragraph from their 2019 report as not describing an actual attack.}
\end{quote}

\section{End-to-end verifiability}

Next to vote secrecy, integrity-related goals form another (and perhaps an even more important) set of requirements on voting systems. Integrity attacks can occur against almost all the steps of voting (e.g. the votes could be altered or deleted, the ballot box could be stuffed with ineligible votes, the tally could be computed incorrectly, etc). Hence, in order to make sure that the end result truly reflects societal preferences, each step should be verifiable. 

The explicit properties to consider, and even a potential technical approach, are first referred to in the 2011 ODIHR report~\cite{ODIHR2011}:

\begin{quote}
    ``In recent years, advances have been made in the field of cryptography to enable end-to-end verification of the votes cast, i.e. a possibility for an individual voter to verify that his/her vote was (i) cast as intended, (ii) recorded as cast, and (iii) counted as recorded. Such individual verifiability usually relies on giving the voter a code that allows him/her to check later whether their vote was correctly recorded or even counted.''
\end{quote}

This is a fascinating statement in several aspects. The first ideas about verifiable online voting can be tracked back to 1980s to the pioneering works of Beanloh \emph{et al.}~\cite{DBLP:conf/focs/CohenF85,DBLP:conf/podc/BenalohY86}. However, by early 2010s, the terminology still hadn't settled, which can be seen from the above quote essentially equating end-to-end (E2E) and individual verifiability. In recent research, it is more customary to follow the verifiability categories by Kremer~\emph{et al.}~\cite{DBLP:conf/esorics/KremerRS10}:
\begin{itemize}
    \item \emph{Individual verifiability}: a voter should be able to check
    that his vote belongs to the ballot box.
    \item \emph{Universal verifiability}: anyone should be able to check
    that the result corresponds to the content of the ballot box.
    \item \emph{Eligibility verifiability}: only eligible voters may vote.
\end{itemize}
For a good overview of different proposed verifiability notions and comparison of their formal definitions we refer to the 2016 paper by Cortier \emph{et al.}~\cite{DBLP:conf/sp/CortierGKMT16}.

Thus, following the Kremer-style approach, only (i) cast-as-intended and (ii) recorded-as-cast checks would fall under individual verifiability, whereas (iii) counted-as-recorded would nowadays be called universal verifiability. 

Whether individual, universal and eligibility verifiability combined give rise to E2E verifiability or not, is still a matter of ongoing academic discussion (and ultimately a matter of definition). For example, the recent report by German BSI~\cite{BSIE2E} gives the following definition:
\begin{quote}
    ``Let $\tau$ be a trust assumption, and let $\mathcal{A}$
be a set of algorithms/attacks that an attacker can execute under the trust assumption $\tau$. We say that
a voting system is end-to-end verifiable under the trust assumption $\tau$ against  $\mathcal{A}$ if under any attack
$A \in\mathcal{A}$ (obeying the trust assumption $\tau$), the probability that the final result published by the voting
system is accepted, even though this result does not match the voters' votes, is negligible.''
\end{quote}

An important feature of this definition is that here E2E verifiability is not a yes/no property, but is parametrized by the trust assumptions and a negligibility threshold. Hence, virtually any voting system is E2E verifiable, but some may need many trust assumptions to be satisfied. Intuitively we may say that such systems are E2E verifiable only in a very weak sense, and we would instead like to have strong verifiability (i.e. as few trust assumptions as possible). 

Unfortunately, the report~\cite{BSIE2E} does not state which level of verifiability can be considered good enough.\footnote{Still, the report thoroughly assesses practicality of various possible approaches to verifiability, hence acknowledging that there is potentially a verifiability-practicality trade-off, where a good balance needs to be found.}\footnote{Also, the scope of the report~\cite{BSIE2E} is limited to studying various verifiability techniques in isolation, and no attempt is made to evaluate more complex systems featuring a combination of these techniques.} This also holds true for the ODIHR reports. 

After the 2011 ODIHR report and emergence of a proof-of-concept malicious voting application~\cite{heiberg2014modeling}, Estonia implemented a solution for cast-as-intended and recorded-as-cast, or simply individual verification~\cite{DBLP:conf/ev/HeibergW14}.\footnote{The Estonian approach to individual verification is now known as cast-and-audit~\cite{BSIE2E}.} This was noted positively in the 2015 report, however, ODIHR rapporteur drew attention to lacking counted-as-recorded (or universal) verifiability~\cite{ODIHR2015}. 

The corresponding mechanisms were added with the 2017 update to the Estonian Internet voting system, code-named IVXV~\cite{DBLP:conf/voteid/HeibergMVW16,EHIN2022101718}. 2019 ODIHR report again took a positive note on the development, but this time the rapporteur gave a remark on insufficient safeguards against insider attackers~\cite{ODIHR2019}. As a motivating example, a scenario involving a voter leaking his/her individual verification QR-code was given. We studied this scenario in detail in Section~\ref{sec:votesecrecy} and concluded that Internet voting is actually more secure than paper voting against this type of attack exactly because it additionally needs a dishonest insider.

The 2023 report~\cite{ODIHR2023} describes another potential scenario.
\begin{quote}
    ``However, it is known that the voter verification mechanism is vulnerable to compromise if the voting client application is altered. While it would be difficult to deploy this attack vector at scale and undetected, the possibility of this type of attack indicates a critical deficiency in the current design of the voter verification step. Namely, the application could be programmed to crash immediately after collecting the information about the properly submitted ballot, but before using this information to conduct the verification. If the voter restarts the application and attempts to vote again, possibly upon incorrectly assuming that their ballot has not submitted, the altered application would then first submit the ballot with an altered choice, but would perform the voter verification procedure on the ballot submitted before the crash, and the voter verification application would display the voter's original choice.''
\end{quote}

The rapporteur is referring to the attack by Pereira~\cite{DBLP:conf/fc/Pereira22} making use of the fact that the current IVXV individual verifiability protocol does not reveal whether the audited ballot was the last one submitted by the voter. This is a conscious design decision taken to minimize the threat of coercion in the scenario where the attacker demands verification of the QR-code to see if it still verifies the vote after some time (attempting to make sure that the voter has not re-voted). 

This is one example of the conflict between verifiability and coercion resistance requirements. Pereira's attack is easy to circumvent by adding the freshness check, but that would increase coercibility to some degree. Is this an acceptable trade-off from the ODIHR point of view?
\begin{quote}
    \emph{We recommend that ODIHR explicitly considers the potential trade-offs implied by implementing its recommendations, and takes an explicit stance concerning which one of the trade-offs is the best.}
\end{quote}

The 2023 report~\cite{ODIHR2023} also proposed an additional auditing step in the universal verification stage. This step was added for the 2024 European Parliament elections. 

However, all these examples bring us to a bigger question mentioned earlier in the section -- what are the criteria under which a voting system can be considered end-to-end verifiable at a sufficient level?

Consider the above-described attack by Pereira~\cite{DBLP:conf/fc/Pereira22} as an example. Is the possibility of such an attack a reason to claim that IVXV does not have (a sufficient level of) E2E verifiability? Let us study the assumptions the attacker has to fulfill. 

First of all, the attacker has to develop and distribute a malicious voting application. The ability to write an application is not a big problem for a mid-level programmer (so no remarkable assumptions here), but distributing and getting it run undetected is not so easy. All the following assumptions need to hold:
\begin{enumerate}
    \item the voter is successfully directed to an unofficial distribution channel,
    \item the voter does not verify authenticity of the voting application, and
    \item the voter does not report suspicious crash of the application.
\end{enumerate}

For assumption 1, it is easy to set up a lookalike website, but it will very probably be noticed and reported. For assumption 2, we need to take into account that under Windows and macOS the voting application is signed with a developer key and the OS verifies the signature before running it. It is possible for an attacker to register as a developer, but this will leave more traces. Linux users are supposed to verify the checksum of the application themselves, but on the other hand, Linux users are more likely to do it. For assumption 3, we note that in order to have a significant effect, the attacker needs to manipulate many votes. However, the probability that no crashes will be reported decreases exponentially fast in the number of crashes (see below for the computations). 

The attack scalability issue is also noted by ODIHR~\cite{ODIHR2023} (``\dots{} it would be difficult to deploy this attack vector at scale and undetected \dots''). But how does this translate to the level of (E2E) verifiability? Let us use the above-cited BSI definition~\cite{BSIE2E} stating ``\dots{}  the probability that the final result published by the voting
system is accepted, even though this result does not match the voters' votes, is negligible.''

What does the ``final result'' mean here? In general, the voting system produces two results -- the voting result (how many votes did every candidate get) and the election result (who got the seats). We argue that the requirement that the voting result has to exactly match the sum of the individual votes is too strong. Most notably, paper voting does not guarantee this as there is human involvement needed to interpret and count the ballots, and this process is a subject to inherent flaws~\cite{goggin2012post}. 

Paper voting operates under the assumption that if there is no large-scale systematic violation (e.g. many polling station workers stuffing the ballot boxes\footurl{https://www.youtube.com/watch?v=5UFIQamKbjg} \footurl{https://www.youtube.com/watch?v=Ep1ef3_wLX0} \footurl{https://www.newsflare.com/video/634198/video-emerges-of-ballot-box-being-stuffed-during-russias-presidential-election} \footurl{https://globalnews.ca/video/4092550/russian-polling-station-workers-accused-of-stuffing-ballot-box-11-times-within-13-minutes/}), the stochastic mistakes cancel out and the election result is a fair reflection of the voters' preferences.

Of course, against large-scale paper voting violations there are safeguards in place. The ballot boxes are publicly observable, so in principle it is possible to at least detect, although not always physically prevent, malicious activities like ballot box stuffing. However, even detection of large scale fraud may be sufficient as it is possible to call the elections void (i.e. not to accept the result) in case the reports of fraudulent actions are frequent.

This is exactly what is happening in case of the Pereira attack in IVXV. Even if the attacker successfully distributes a malicious voting application and gets the voters to run it, they still must notice the crash. Some percentage of the voters will also report it, and this way the attack gets detected by the election organizer. The matter can be studied and, in case the reports are frequent, the election result may get rejected.

If there are $n$ voters who run the manipulated application, and a voter has probability $p>0$ to report the crash, then the probability that the attach remains unreported is $(1-p)^n$ which converges to $0$ exponentially fast in $n$. Thus, the probability that the manipulated result gets accepted undetected is negligible\footnote{We say that a function $\mu:\N\to[0;1]$ is negligible if for every positive polynomial $p$ there exists $N\in\N$ such that for every $n>N$, the inequality $\mu(n)<\frac{1}{p(n)}$ holds. It can be shown that for any $q\in[0;1)$, the function $q^n$ is negligible.}. 

We conclude that Pereira's attack, even though possible in principle, does not necessarily violate E2E verifiability (at least the BSI flavour of it). Of course, there are more details to consider, e.g. how does the reporting frequency affect the above probability computation, and how frequent do the reports have to be to get the election result rejected. The latter question is also raised in the 2011 ODIHR report~\cite{ODIHR2011}:
\begin{quote}
    ``Although the Election Act indicates that the NEC can invalidate the results of the Internet voting, it does not specify on which basis and under which circumstances the results of the Internet voting can be declared invalid. It further does not specify how and by which means voters can be informed that they have to recast their vote on paper on election day.\\[1ex] 
    The OSCE/ODIHR recommends that legal provisions with regards to all stages of the Internet voting, including conditions for invalidation of the Internet voting results, are further detailed and consolidated in the law.''
\end{quote}

The intent of this recommendation is good, but it is not clear what is the correct level of technical detail such conditions should be written into the law. Invalidation of (part of) the voting results is a very prominent action, and the state may want to retain the discretion to decide flexibly based on the concrete circumstances that are hard to foresee in the time of writing the legislation. Thus, in 2017 an amendment in the following wording entered force\footurl{https://www.riigiteataja.ee/en/eli/ee/510032014001/consolide/current}:
\begin{quote}
    ``\dots{} the National Electoral Committee has the right \dots{} to annul the votes cast in the advance voting in part or in whole due to material violation of law and call on the voters to vote again during advance voting or on the election day;''
\end{quote}
Thus, the flexibility to consider the circumstances is there. On the other hand this of course means that it is impossible to give a strict mathematical proof that depends on the exact conditions for annulling the votes.\footnote{Extending the argument we can see that this problem is inherent. The target of verification is detecting anomalies. In case of elections, the anomalies can hardly be fixed, so the only option is to cancel the voting event and try again. A precise mathematical definition of E2E verifiability hence has to rely on the (mathematical) criteria for annulling the result. In the real life, however, the annulling decision is taken by the lawyers who prefer not to be pre-committed to strict criteria, but rather have the power of discretion. It is ultimately the question of who gets to decide upon annulling the elections -- mathematicians or lawyers. The current societal order seems to favour the lawyers in this role. Thus the mathematicians have to come up with E2E verifiability definitions that allow for the power of discretion for the lawyers, or to acknowledge that E2E verifiability can not be achieved in practice at all.}

However, these details do not change our main message -- just the existence of some attacks does not automatically mean that the system is not E2E verifiable on a sufficient level.
 
This brings us to the next important question -- how does one exactly decide whether a given voting system is (sufficiently) E2E verifiable? We have been unable to identify an existing  good methodology for that in the literature. The BSI report~\cite{BSIE2E} explicitly leaves complete systems out of scope, and the recent ODIHR handbook on observing ICT in elections~\cite{ODIHR_ICT-Handbook} does not present any methodology either. 

However, we argue that such a methodology is urgently needed. If ODIHR never actually says that the system is good enough, and only keeps giving recommendations on issues that actually do not violate E2E verifiability, it becomes increasingly hard for the election organizers to do their job. 

First, if there are no clear criteria that ODIHR uses to decide upon the level of E2E verifiability, it is unclear what should be the target of development. If the ODIHR rapporteur keeps pointing to marginal issues (and one can always find something to point to), the target keeps moving forever. Of course, the environment is changing, and the criteria should sometimes also change to follow the environment, but there should explicit and clearly communicated criteria at any given point in time. Moreover, these criteria should be communicated well ahead of the assessment in order to give sufficient time for their implementation.

And second, at least in the case of Estonia, any recommendations given by ODIHR get interpreted in the conservative media as evidence to support the narrative that electronic voting is insecure. Without the explicit claims in ODIHR reports referring to acceptable levels of verifiability, the election organizer is subjected to undue pressure from the critics. But of course, in order to give such claims, ODIHR first needs to create a methodology how such claims can be reached.

\begin{quote}
    \emph{We recommend that ODIHR develops and publishes clear criteria according to which it decides about (i) the level of verifiability of the studied voting system, and (ii) whether this level is sufficient or not.}
\end{quote}

\section{Recommendations that do not help}

With 41 recommendations, it becomes statistically likely that there are some among them that seem quite good at the first glance, but are impossible to implement or contradict some other requirements. In this Section we review some of the more interesting examples of this category.

The 2011 ODIHR report~\cite{ODIHR2011} notes and recommends the following.

\begin{quote}
    ``Daily update of the voter register during the voting period as required by the Election Act was performed together with the daily backup of data. The project manager accessed the servers for daily data maintenance and backup breaking the security seals and using a data storage medium employed also for other purposes. This practice could potentially have admitted the undetected intrusion of viruses and malicious software. \\[1ex]
    It is recommended that no maintenance of the Internet voting system servers is performed from the start to the end of the Internet voting process.''
\end{quote}

Using a single-purpose data transfer medium is indeed a good suggestion, but otherwise the recommendation to avoid maintenance violates the best practices of running IT-systems. Daily backups of the digital ballot box are essential to ensure continuity of elections even in the case of a major disruption. Also, daily updates of the voter register are required by the Election Act for a reason as the list of eligible voters changes every day (including in an unpredictable manner due to deaths). Thus, the running system needs to be accessed every day, and physical access is actually easier to secure. Implementing a remote console access for these maintenance tasks would potentially be a considerably larger security vulnerability as it would be easier to misuse without detection.

\begin{quote}
    \emph{We recommend that ODIHR aligns its recommendations with the best practices of maintaining IT systems, and considers alternatives before selecting its recommendations.}
\end{quote}
The 2023 ODIHR report~\cite{ODIHR2023} reads:

\begin{quote}
``During the election period, internet voting continued to enjoy a high level of public trust, owing to the transparency of the system and this voting method being an established practice. However, some voters distrust the results of internet voting, with notable divisions within the society between those who fully trust and those who fully distrust internet voting. This was also exemplified by the difference in the preferred political forces of those who voted in polling stations and those who voted online.  This polarization was also represented in the political spectrum, with some parties resolutely supporting internet voting and other parties raising doubts before and after the elections, most notably EKRE. Following the announcement of results, some EKRE frontrunners, including its leader, made statements that the elections were stolen through internet voting and the party subsequently submitted several complaints requesting the annulment of internet voting, which were all dismissed. Notably, the allegations on systematic electoral fraud were made in the public domain without substantiation; such claims can harm public trust in democratic institutions. While the election authorities provided comprehensive voter information on its website and other platforms, they did not timely and proactively address some of the concerns raised by the contestants regarding the integrity of the internet vote.\\[1ex]
To further increase and maintain trust in internet voting, the election authorities should proactively address all concerns raised by election stakeholders who distrust the results of internet voting.''
\end{quote}
A similar recommendation was also given in 2019~\cite{ODIHR2019}:
\begin{quote}
    ``Other significant risks that may negatively affect public confidence in Internet voting include cyber-attack allegations from disinformation campaigns or human error.\\[1ex]
    The SEO could review the potential effects of cyber-attack allegations against the Internet voting infrastructure, and develop a risk mitigation strategy.''
\end{quote}

These recommendations are impossible to implement in practice. First, the critics of Internet voting have shown remarkable creativity in coming up with a wide spectrum of claims, and addressing all of them proactively is impossible. Second, it is important to realize that the claims and allegations against Internet voting are not motivated by true concerns of security (in which case argumentation would be possible). Rather the goal of the opposing politicians seems to be discontinuing Internet voting altogether, hoping that younger and more liberal part of the electorate would then refrain from voting at all. One can not possibly address this goal via proactive communication.

\begin{quote}
    \emph{We recommend that ODIHR should not give recommendations that are impossible to fulfill.}
\end{quote}
The 2023 report~\cite{ODIHR2023} recommends:

\begin{quote}
    ``At the beginning of internet voting, on 27 February, the system was configured with an outdated voter register, which resulted in 63 voters casting votes for lists in their previous districts of registration before the misconfiguration was discovered and corrected. The voters were informed, and 59 of them voted again online, and one also voted at a polling station. The ballots of the four voters who did not vote again were revoked by the SEO, effectively invalidating their votes. The lack of quality assurance of configuration and other data can negatively affect the voters' trust in internet voting. \\[1ex]
    To prevent errors or outdated information when configuring the components of the system, the election authorities should put in place a quality assurance process that includes the comprehensive testing of the internet voting system in its entirety before being deployed.''
\end{quote}

This was not an issue of the voting system misconfiguration, but rather a problem of the population registry not having updated information about the people who had recently moved to another electoral district. This problem would have been impossible to detect with any quality assurance on the Internet voting system (even considering the population registry as a part of the ``Internet voting system in its entirety'').  Rather it is worth noting how quickly it was possible to fix the problem, and the very limited impact it had as a result of the fast reaction.

\begin{quote}
    \emph{We recommend ODIHR to acknowledge that not all the potential problems can be prevented. For some issues it is more reasonable to detect the problem, and have a team ready to act promptly on it.}
\end{quote}

\section{Conclusions}

ODIHR has done an impressive amount of work to raise the standards of electoral integrity amongst its member states, but also on a larger international scale. Both its observation reports and handbooks provide a valuable source for all the countries interested in improving their democratic processes.

Since introduction of Internet voting in Estonia in 2005, ODIHR has consistently reported on its state of development and given recommendations for improvement. Many of the recommendations have proven very useful and have been implemented in the legislation and/or in the system itself. Some, however, require further discussion and analysis. 

ODIHR has relatively limited options for disputing or retracting its recommendations. There may be a good reason behind that -- if raising a dispute would be too easy, ODIHR would need to put a lot of effort and valuable resources into dispute resolution.

On the other hand, if some recommendation is impossible to implement or significantly contradicts other requirements, it may end up causing more harm than good. Mere existence of unimplemented recommendations in ODIHR reports can be used to raise ungrounded allegations against the security of Internet voting.

Due to the lack of a better option, the author decided to share some of his thoughts in the form of an academic paper. The paper has also been deliberately formatted in the style of ODIHR reports.

The author hopes that the presented thoughts are useful for ODIHR (and other similar organizations) to improve their election observation processes, especially concerning Internet voting. 

And, to remain true to the ODIHR style, let us conclude the paper with one final recommendation.

\begin{quote}
    \emph{We recommend that ODIHR makes discussing and disputing its recommendations more accessible. Also, clear process and criteria for retracting the recommendations should be developed.}
\end{quote}

\subsection*{Acknowledgments}

The paper has been supported by the Estonian Research Council under the grant
number PRG2177.

\bibliographystyle{splncs04.bst}
\bibliography{recommendations}

\end{document}